\def\arcs{$^{\prime\prime}$~}
\def\et{{\it et al.~}}
\def\page{\vfill\eject}

\def\gtorder{\mathrel{\raise.3ex\hbox{$>$}\mkern-14mu
             \lower0.6ex\hbox{$\sim$}}}
\def\ltorder{\mathrel{\raise.3ex\hbox{$<$}\mkern-14mu
             \lower0.6ex\hbox{$\sim$}}}
\def\asec{^{\prime\prime}}
\def\Rf{\normalbaselines\parindent=0pt \medskip\hangindent=3pc \hangafter=1 }

\documentstyle[12pt,aaspp]{article}
\begin{document}
\large
\centerline
{M51 Stripped To Its Bones}
\normalsize
\vskip 1cm
\centerline{Hans-Walter Rix\footnote{Hubble Fellow}}
\centerline{\it Institute for Advanced Study, Princeton, NJ 08540}
\centerline{and}
\centerline{Marcia J. Rieke}
\centerline{\it Steward Observatory,}
\centerline{\it University of Arizona, Tucson, AZ 85721}
\vskip 2.0in
\centerline{Submitted to {\it Ap.~J.}, March 2, 1993}
\par\noindent
\page

\abstract{

We present optical and IR surface photometry of M51 (NGC~5194) at
B~V~R~I~J~K and CO$(2.3\mu )$.
These data are used to establish whether K-band ($2.2\mu$) images
of spiral galaxies provide reliable maps of stellar surface mass
density features such as massive spiral arms or bars.
The main distorting agents in the mapping at shorter wavelengths are
dust extinction and luminous young stars.
{}From modeling the color changes across the main dust lanes in M51, we
find the optical depths to be $\sim0.5$ in the K-band. For these optical depths
the
K-band flux is attenuated by only $\ltorder 10\%$
even in the dust lanes.
{}From monitoring the gravity-sensitive CO$(2.3\mu )$ index across the
spiral arms we find that young, red supergiants do not distort significantly
the K-band image except in one small patch.
OB associations are visible in the K band images but
only cover a very small fraction of the spiral arms.
On this basis, {\it we conclude that
K band images do
trace the massive disk star population and allow
a mapping of the azimuthal variation in the
surface mass density of the stellar disk}.
In M51 we find the surface mass density contrast (arm/inter-arm)
to range from 1.8 to 3, comparable to results from N-body simulations
of the galaxy's tidal encounter with NGC~5195. This density contrast is {\it
larger} than the light contrast
in I band images, where the spiral arm crest is affected by dust extinction.
The spiral arm amplitude in M51 clearly shows smooth, strong radial
variations, with a maximum at $\sim 130\asec$ and minima at $45\asec$ and
$170\asec$. These variations may arise from interference of a pre-existing
spiral pattern with the tidally induced spiral arms.
An ongoing K-band imaging study of a sample of spiral galaxies will yield a
more
representative picture of the role of bars and massive spiral arm features.

}

\smallskip\par
\noindent{\it Subject Headings: } galaxies:~stellar
content,spiral,individual:~M51 --

infrared:~galaxies -- ISM:~dust,extinction

\page

\section{INTRODUCTION}

The surface brightness in elliptical galaxies
appears to map the stellar surface mass density very well (except for very
shallow
radial gradients in the M/L). In contrast, it is
difficult to map variations in the stellar surface mass density in spiral
galaxies. The mapping is complicated both by the prominent dust lanes and
patches in many spirals and by young stars which may contribute much to the
light
in many colors but only little to the total stellar mass (e.g Schweizer 1976).
Mapping the azimuthal surface mass density variations in spiral galaxies is
important in several respects:
first, it allows one to make an unbiased estimate of the frequency of
bars in spiral galaxies. Some bars only become apparent in the infrared.
Second, mapping the position and amplitude
of massive spiral arms is important both for testing density wave theory
and for comparing dynamical simulations of tidally induced spiral arms
(Toomre and Toomre 1972, Hernquist 1990) directly to observations.
Furthermore, this mapping allows us to check whether galaxies with flocculent
or chaotic spiral arms (in young stars) have any organized underlying
density fluctuation.
Finally, we can see how the spiral arm mass amplitude in isolated
spiral galaxies correlates with the current star formation rate,
as expected if the spiral structure is driven by star formation.

In this paper we employ optical and IR surface photometry of M51
to establish firmly that K band imaging with IR arrays is
a reliable and efficient way to map surface {\it mass} variations\footnote{
The light in the K band is of course dominated by giants, which make up only a
small fraction of the stellar mass. However, for an old population
giant stars have the same
spatial distribution as the main sequence stars}
through surface {\it brightness} variations.
To this end we will show that neither dust nor young, luminous, red stars
strongly affect the K band image, so that it indeed traces the
old massive disk population.

We use M51 for this initial effort because it is a nearby galaxy, yet not too
large
to be imaged, and because it is the prototype of a grand design spiral.
Furthermore, previous observations have shown tentative evidence of a central
bar which our data can confirm.
It is important to keep in mind that, due to the recent tidal encounter
with NGC5195, M51 is in an atypical dynamical state, and hence conclusions
about the properties of its spiral arms must not be generalized.

Aside from these scientific questions we also will address
some more technical aspects of infrared surface photometry.
No body of near-infrared surface photometry which could be used to define the
typical run of colors and surface brightnesses currently exist. Most
studies to date have concentrated on starburst or Seyfert galaxies with
the exception of the study by Peletier and Willner (1991)
of Virgo spirals and of the
study by Wainscoat \et 1990 of edge-on galaxies where the surface brightnesses
are much higher than in face-on objects.
Therefore it appears worthwile to examine the accuracy,
and limitations, of such data sets.

The remainder of the paper is organized as follows:
in Section 3 we describe the observations and data reduction
procedures; in Section 4 we analyze and model the data
to address the questions raised above and in Section 5 we
summarize and discuss the results.

\medskip
\section{OBSERVATIONS AND DATA REDUCTION}
\smallskip\par
The data collected for the present work are comprised
of ``optical" CCD data (in B,V,R,I and H$\alpha$) and
IR data (J,K and CO$2.3\mu$) acquired with NICMOS2
and NICMOS3 arrays. Since the data reduction procedures are
quite different for the two sets we will describe them separately in the
following
paragraphs.

\smallskip
\subsection{CCD Data}

The data were obtained on September 4, 1989 at the Steward 2.3m telescope on
Kitt Peak
with a focal reducer,
using a TI CCD, with $800\times800$ pixels of $0.55$\arcs each and
a field of view of $7.3^{\prime}\times7.3^{\prime}$. The exposure times
for the images were 90 seconds for V,R and I, 120 seconds for B
and 300 sec for H$\alpha$. Flat-fielding was done with dome and twilight flats
and is good to $\sim 1\%$ over the inner $5\times 5$ arcminutes.
The images were calibrated using CCD standard fields in M67 (Schild
1983).
The uncertainty in the zero points of the photometry in the various bands
is estimated to be $0.05$mag
from internal comparisons and from comparison with Schweitzer (1976)
and Boroson \et (1981).
However, the shape of the luminosity
and color profiles presented
below depend only on the linearity of the array and the sky subtraction
and are therefore constrained better than the zero point at small
radii ($4^{\prime\prime}\ltorder R\ltorder 100^{\prime\prime}$).
For an uncertainty in the sky level of about 0.5\% (variance of sky level
in four ``empty" patches) the error in the luminosity profiles exceeds
10\% outside of 150\arcs .

As a last step in the reduction the CCD images were rotated and rebinned
to the K band image of lower spatial resolution (see below).
Images in all bands are shown in Figures 1 and 2.

\smallskip
\subsection{IR Data }

The data for the K band mosaic of M51, shown in Figure 1f, were obtained
on April 15, 1989 at the Steward 1.5m telescope on Mount Bigelow.
The detector was a $128\times128$ NICMOS2 array, with a scale of $1.8$\arcs
/pixel.
One frame of the mosaic was centered on the nucleus of M51, while the
other four frames had the nucleus about 20 pixels from each corner
of the array; the final mosaic encompassed an area of
about $7.2^{\prime}\times7.3^{\prime}$.

At $2.2\mu$ the sky is the dominant signal at virtually  all points
of the image. Therefore, careful monitoring of the sky is mandatory. To this
end
the telescope was wobbled after each 60 second exposure on the target
to a patch of sky off-set by about $10^{\prime}$ where an equally long
sky exposure was taken. The total integration time (on the object)
was about 20 minutes per mosaic patch. The object-sky data stack was reduced as
follows
(see also Rix 1991, for more details): each frame was bias subtracted and
all bad pixels (covering $<0.5\%$ of the total area ) were replaced by the
median
value of the surrounding eight pixels. Then, each object frame was divided
by the average of the two bracketing sky frames and the mean
count level of the two sky frames was subtracted. These procedures result
in a stack of typically 10 to 20 sky-subtracted and flat-fielded
object frames. The images in this stack are then
aligned spatially, using point sources in the frames.
Finally, the five mosaic pieces were aligned and combined into a single image
using software kindly provided by R. Guhartakurta and G. Bernstein.
To make a first order correction for any tilts in the final image,
arising from the mosaicing, we fitted a plane to
the same patches of sky which had no detectable galaxy signal in the CCD frames
and subtracted it. The final picture is shown in Figure 1f.

The flux calibration of the image was achieved by integrating the
sky-subtracted flux in the
image in concentric annuli and comparing the result to the
aperture photometry by Aaronson (1977). We estimate the K zero point to be good
to $0.05$mag
(Figure 3).

Further data at J$(1.3\mu )$, K$(2.2\mu )$ and CO$(2.3\mu )$ were obtained at
the Steward 2.3m
telescope using a NICMOS3 array on January 23, 1992.
These data have much higher resolution:
0.6\arcs~/pixel (with 1.2\arcs seeing), but only cover an area of
$2.4^{\prime}\times 2.4^{\prime}$. These data were reduced and calibrated
as described above, and the K band data provide an excellent external accuracy
check, since they were taken at a different spatial resolution, with a
different
detector on a different telescope. These images are shown in Figure 2a-c.
Again, aperture photometry from Aaronson (1975) was used to flux calibrate the
J and K frames. The CO($2.3\mu$) filter used in this present work
deviates from the standard
passband (Frogel \et 1978) and we will only use it for differential
spatial measurements.

\subsection{Photometric Profiles and Colors}

We present the photometric profiles in all colors as the surface brightness
averaged in concentric annuli as a function of the mean radius
of the annulus, rather than fitting ellipses to the image.
This is done because M51 is seen nearly face-on and because
the ellipse fitting would be dominated by the spiral arms over
a significant range in radii. All strong point sources have
been masked out in the profiles shown here.
The global light profile is insensitive
to the details of the point source removal, even in B.

Figure 4 shows the profiles in all colors. The J band data and
some of the K band data
extend over a smaller radial range because of the smaller area covered
by the NICMOS3 data. The K band data in the overlapping region
may be used to judge the quality of the K band surface photometry.
These data show that even in the infrared useful data can be obtained
out to radii of $\sim 150\asec$.
The ``shoulder"  in the light profile at $\sim 15\asec$ arises
from the small central bar in M51 which becomes very clear in the infrared;
its properties will be discussed elsewhere (Zaritsky et al. 1993).

Figure 5 shows the radial run of the optical and IR colors.
The flatness of the R--I profile ($\Delta(I-R)\ltorder 0.03$mag), which should
be insensitive
to population changes and dust (see Section 4.1.2), can be used to check the
reliability
of the optical color profiles.

\medskip
\section{ANALYSIS}
\smallskip\par
\subsection{Global Colors and Color Gradients}
\bigskip
\subsubsection{Comparison with Previous IR Aperture Photometry}

Except for a few cases of edge-on galaxies (Wainscoat \et 1990) and
Virgo spirals (Peletier and Willner 1992) the available photometric information
in the IR on disk galaxies stems from aperture photometry
(Aaronson 1977, Bothun \et 1985).
The good agreement between the integrated surface photometry
and aperture photometry (Section 2.2), along with the smoothness of the $J-K$
color profile (Fig. 5) and the IR surface photometry of ellipticals
(Peletier 1989 and Rix 1991), reaffirms the quantitative reliability of
IR surface photometry with detector arrays.
The globally averaged  J-K color, 0.95mag, and an
estimate of B-H\footnote{This estimate assumes that $H =
J + 0.3(K-J)$}, 3.5 mag, agree well with values found by
Bothun \et (1985) for more distant spiral
galaxies of an absolute magnitude in B of $\sim -20.5$.
\bigskip
\subsubsection{Dust and Global Color Changes}

It is tempting to use this data set
to test claims
made in a flurry of recent papers (e.g. Disney \et 1989,
Valentijn 1990) that the disks of spiral galaxies
are opaque at optical wavelengths due to dust extinction.
If this were the case then the optical and IR colors should be affected
by such dust. Disney \et (1989) have stressed that the color
changes for a given optical depth may be smaller than expected
for the usual `source -- dust screen -- observer' geometry, because
the dust is mixed in with the stars. Considerable
amounts of dust may therefore cause only small color changes.

Virtually all previous authors have only accounted for {\it absorption}
when quantifying the optical depths which are consistent with
the observed colors. Dust {\it scattering} has always been neglected,
with the notable exception\footnote{Witt \et (1992) and
Bruzual \et (1988) included scattering into their calculations
but did not discuss the applicable geometry of plane parallel
stratifications of dust and stars with different vertical scale heights}
of D. Elmegreen (1980).
Since the (isotropic) albedo of typical dust grains (see e.g. Draine and Lee
1984)
in the optical is about 0.4, almost as many photons are reflected
as are absorbed. Thus a dust layer in the disk plane does not only
act as a sink for the light behind it, but also as a mirror
for the light in front of it. Using the simplified radiative
transfer model described in Section 3.2 and in the Appendix, we can illustrate
how the colors of M51 (assumed to be seen at cos(i)=0.9)
change with increasing optical depth, assuming the dust
has a scale height equal to 0.4 times that of the stars. The relative scale
height was chosen to be comparable to those found for the dust lanes in M51
(see Section 3.2). Figure 6
shows the change relative to the dust free case for all the
colors shown in Figure 5. Note that for small optical depths
the optical fluxes are slightly enhanced compared to K, where
scattering is less important. Only for optical depths
larger than $\tau_V=1$, do the color changes become larger
than the observed gradients within M51. Thus large optical
depths $\tau_V>>1$ can be ruled out on the basis of this data
set (in conflict with Valentijn's claim), but no tight limits
on $\tau_V$ can be derived. A mean optical depth of $\langle
\tau_V\rangle\sim 0.5-1$ is consistent with the data.
In turn, Figure 6 illustrates that for face-on disks the observed global colors
are insensitive to the dust content ($\Delta \ltorder 0.05$) for optical
depths of $\tau\ltorder 0.5$.

\bigskip
\subsubsection{Optical-IR Color Gradient}

It is apparent from Figure 5 that the bulge and the disk
are very similar in colors redward of V.
However, the disk exhibits a clear color gradient in the V and B colors
of $d(I-V)/dln(r) = 0.3$ and  $d(I-B)/dln(r) = 0.5$, respectively.
Within the bulge
the B and V gradients are similar to the disk's,  but there is
a color off-set towards the red
of $0.3{\rm mag}$ in B and of $0.15{\rm mag}$ in V.

The optical and IR color gradients of the bulge are
worth being considered separately.
Elliptical galaxies show relatively homogeneous properties
in their optical to IR color gradients, $d(B-V)/d(V-K)=0.25\pm0.08$
(Peletier 1989). The gradients in ellipticals are thought to arise
principally as a result of a gradient in the metallicity. The attribution
of these gradients to metallicity is also supported by the variation of B-V
with V-K, $0.33\pm0.03$, for a sample of field galaxies observed
by Persson \et (1979). This variation is between galaxies referenced to a
fiducial fraction of the galaxy's light, and color variations between these
galaxies result from the metallicity differences which are correlated with
the luminosities and hence masses of the galaxies. We find a color gradient in
the bulge of M51
of $0.51\pm0.08$, i.e. $V-K$ changes less than expected from the
$B-V$ gradient.

We consider three possible causes for the size of M51's color
gradient. First, it could arise from an extinction gradient with more dust
covering
the inner portion of the bulge than the outer portion. A simple screen with
$A_V\approx0.75$ at the center and declining to zero at 20" would reproduce the
V-K gradient but could not simultaneously reproduce the gradients observed in
the
other colors. Inspection of Figure 6 indicates that more complicated dust
models
are not likely to produce the $d(B-V)/d(V-K)$ observed for M51's bulge.
Second, M51's gradient could result from a change in the metallicity.
Comparison with the gradients for ellipticals mentioned above suggests that
this
explanation may account for some portion of the gradient. The third possibility
would account for the gradient through an outward decrease in the mean age of
the
stellar population. We examined the magnitude of the effect possible from an
age
change using models similar to those calculated in Rieke \et (1993).  A slope
of
$d(B-V)/d(V-K)=0.39$ is derived for old galaxies with the V-K gradient in M51's
bulge corresponding to an age change of about a Gyr. Again, this explanation
cannot alone account for the ratio of optical to infrared color gradient
observed
in M51's bulge. We cannot separate age effects from metallicity effects with
our
limited set of colors, but the correct explanation for the gradients may well
involve a combination of the two.

It must be noted that these statements exclude the mini-AGN region in the
nucleus.
The region with $r<2\asec$ has colors substantially bluer than the surrounding
bulge.

\vskip 0.8truein
\subsection{The Structure of the Dust Lanes}
\medskip
The apparent structure
of M51 in the optical differs from its stellar surface mass distribution, at
least in part,
through the light obscuration in dust
lanes and patches (Figure 1).
Before identifying variations in the K band image with surface mass variations
we must assess the impact of those dust lanes at K.
The structure of dust features in a sample of spiral galaxies has been
studied carefully by Elmegreen (1980), based on optical photometry. She found
that
the optical depths in the lanes can be large, $\tau_V\approx 10$.
The fact that the dust lanes and many dust patches appear so prominently
in I band images (Fig. 1) implies that
even at $8000$\AA~ the optical depths in these regions are large.
However, in the K band the dust lanes appear to be optically thin from Figures
1 and 2.

Aside from Elmegreen's study (1980), there is very little
information on the vertical geometry of the dust in
galaxies outside the Milky Way.
Kylafis and Bahcall (1987) and Wainscoat \et (1990)
made an estimate of the global scale height of the dust,
relative to the stars,
from the photometry of edge on galaxies. However, such
estimates refer to the combined effects
of smoothly distributed dust, diffuse dust, dust lanes,
globules and warps in the disk.
Since M51 is nearly face-on it allows
local estimates of the vertical dust distribution.

Figure 7 shows the color changes of all other bands compared to K
along two cuts
perpendicularly across major dust lanes.
These cuts show that the dust lanes are optically thin at K,
semi-transparent at J and I, and optically thick
at B,V and R. We can make a reasonable
estimate of the the dust lanes' optical depths and $z$-heights
with the simple model described in the Appendix.
The models are parametrized by the optical depth at K, $\tau_K$,
and by the relative scale height of dust and stars, $\zeta$.
We assume that the optical depth
outside the dust lane is negligible
and that the unreddened colors do not change
across the dust lane. The latter is supported by the
similarity of the colors on either side of the dust lane,
compared to the color changes across it (see Figure 7).

The adopted color changes in BVRIJ across the
two lanes are listed in Table 1,
along with estimates of their uncertainty.
The quality of match between the radiative transfer models and the observations
is
given by
$$\chi^2(\tau_K,\zeta )= \sum_{all\  colors\  i}{ {[\Delta
C_i^{pred}(\tau_K,\zeta )
-\Delta C_i^{obs}]^2}\over{\delta C_i^2} } .$$
where $\Delta C_i$ is the color difference, $i-K$,
between the center of the lane and the surrounding
innerarm regions and $\delta C_i$ is the error in this estimate.
Contours of $\Delta\chi^2\equiv \chi^2(\tau_K,\zeta
)-\chi^2_{min}(\tau_K,\zeta)$ are shown in Figure 8 for the range
$0<\zeta<1$ and $0<\tau_K<1$.

The two examples show that optical depths in the dust lanes at K are about
$0.5$, corresponding to
$\tau_V\sim 4$. These estimates are in agreement with
Elmegreen's (1980) results. They also agree well with the optical depths
expected
from recent, high resolution CO column density measurements
in the spiral arms of M51 (Rand and Kulkarni 1990),
assuming a standard
$CO$ to dust conversion.
At these optical depths the K band flux is attenuated only by $\sim 10\%$.

{}From analyzing several patches ({\it e.g.} Figure 8),
the relative scale height of the dust to the stars, $\zeta$, is found to range
from
$0.25$ to $0.6$, decreasing with radius. Near the bulge, the dust's vertical
distribution
is comparable to the stars, while it is substantially flatter in the disk.
However, in all cases the dust in the lanes appear to be more vertically
extended than inferred
from global edge-on estimates, $\langle \zeta \rangle = 0.2$.

\bigskip
\subsection{The Structure of the Spiral Arms}
\bigskip
One of the long-standing  problems for testing
models of spiral structure in detail has been the difficulty in mapping
surface mass variations quantitatively. These variations should be reflected
in smooth surface brightness variations of the old stellar population.
However in optical images of spirals, the most conspicuous variations
are caused by dust lanes and by young stars. Even though Schweitzer's (1976)
work has shown that there is a smooth underlying
variation in the red light, its quantitative assessment has proven difficult,
because even images at I are afflicted by dust extinction and by
population changes across the spiral arms.
The uncertainties about the basic structural parameters are
best illustrated by the conflicting claims found in the literature:

\noindent$\bullet$ Schweizer (1976) estimates an arm/interarm contrast of
$\pm 20\%$ at the radii of the outermost HII regions
($\approx 250$\arcs ), while Elmegreen \et (1989)
estimate the contrast to be more
than two magnitudes at the same radii.

\noindent$\bullet$ Elmegreen \et (1989) claim that
the B and I spiral arm amplitudes
are comparable, indicating a density wave rather
than a population change,
while Wright \et (1991) claim that the K band
amplitude is a factor of two lower
than at I, implying that not even the
I band light traces the old disk population.

\subsubsection{Localized Optical-IR Comparison of the Spiral Arms}
\medskip
As a starting point we compare the I band and K band
properties of the spiral arms, because the I band
has been conventionally used (Schweizer 1976,
Elmegreen \et 1989, Kaufman \et 1989)
to assess the dynamical structure of spiral galaxies.

A visual comparison of the I and K band images in Figure 1
shows that many discrete clumps and point sources
(mostly OB associations) are nearly as prominent in K as they are at I;
indeed a few appear only at K because they are completely obliterated
by dust at I. However, these clumps occupy only
a very small fraction of the spiral arms.
The most dramatic difference between I and K is in the appearance of the dust
lanes:
while at I the dust lanes are still prominent (reducing the observed peak flux
to nearly half in some
places), they are discernible at K only just outside the bar (at
$r\approx15$\arcs ).

The differences are highlighted in the J-K color map, shown in Figure 2d.
A more quantitative impression of these color differences, $\Delta (I-K)$ of up
to $0.5$,
is given in two cuts perpendicularly across the spiral arm shown in Figure
7. Thus the local structure of the spiral arm (e.g. the location and the
amplitude of the
density maximum) is inadequately described by the I band data (see also Figure
12).

\subsubsection{How Much Do Young, Red Stars Contaminate The K Band Image~?}
\medskip
Aside from dust obscuration, the spiral amplitude of the light may
differ from the mass through contamination by young stars. In addition to the
blue hot
young stars, cool, red supergiants (RSGs) appear in a young stellar population
after 5 million years.  These stars emit most of their
light in the near infra-red, and could contribute a significant fraction of
the K light in the spiral arms and hence could spoil the K-band light as a
tracer
of the surface density variation.
The relevant definition of ``young" here implies
that the stars were born near the crest of the spiral wave
and have not yet had time to drift away from spiral arm.
The radial distance by which a star will have moved from its
birth position, relative to the spiral density pattern, can be estimated as:
$$\Delta R \sim \Omega_p ~ R ~ T ~ sin(i),$$
where R the distance from the galaxy center, $\Omega_p$ is the pattern speed at
R,
$T$ the age of the star and ${\rm sin}(i)$ is the pitch angle of the spiral
arm.

In a single age population (e.g. formed
by the density wave shock), RSGs contribute
most to the K light after about $1.5\times10^7$yrs
(Rieke \et , 1993). At a $R=25$\arcs,
or about $1$kpc, in M51
this translates into a maximum supergiant contribution at
$\Delta R=9$\arcs outside the dust lane (their approximate birth place),
where we have assumed
$\Omega_p\approx 60km/s/kpc$ and ${\rm sin}(i)=15^{\circ}$ (Tully 1974).
Thus if, RSGs contribute to the K band spiral amplitude
they are expected to do so significantly further away from the dust lane
than the OB associations. Since these stars also had more time to disperse
they will not appear as tightly clumped as the OB associations.

Arm inter-arm comparison of the strength of
CO molecular absorption at CO($2.3\mu$), which is strongest in cool supergiants
with low surface gravities, enables us to estimate observationally
the contribution of RSGs to the K light near the spiral arms (Figure 9).
This test for population changes has the great advantage that it compares
fluxes at nearly identical effective wavelengths and is therefore independent
of dust extinction.
For a single aged population $K-CO$ can differ by $0.1-0.15$mag compared
to an aged ($T>>10^8$yrs) population of the same
metallicity (Rieke \et 1993).
Figure 9 shows the K and CO profiles
across the inner spiral
arms at two positions,
spanning the extremes found in the data set.
In the one arm near the bulge $K-CO$ does not change
measurably across the arm, $\Delta (K-CO) \ltorder 0.025$mag, thus at most
20\% of the light could arise from supergiants. In the second arm $\Delta
(K-CO) \sim 0.1$
over a small localized region, indicating that much of the light in this patch
comes from RSG.
The region of strong CO is about $6\asec$ outside the density crest of the
spiral arm, consistent
with the hypothesis that these RSG's were born just inside the crest 15 million
years ago.
However, significant changes of K-CO  were found only in one single, localized
region.
Thus at most small portions of the spiral arm at K are afflicted by young star
light,
while the rest traces an old population.

\subsubsection{Harmonic Decomposition of the Spiral
Pattern and Global Amplitudes}
\medskip
The most straightforward way of presenting quantitative information about
the spiral arms is to decompose the disk light into its harmonic, or Fourier,
components in concentric annuli, after having corrected for the inclination
of the galaxy (e.g. Grosbol 1987, Elmegreen \et 1989, 1992).

Following Tully (1974) we assumed an inclination of $68^{\circ }$, to transform
the images to face-on.
For radii between 20\arcs and 210\arcs
we formed 30 azimuthal bins
by medianizing the pixel values within segments of $12^{\circ}$ in angular
extent and of
$R_{max}/R_{min}=1.05$ in radial extent.
Taking the median over the sampling area reduces
the impact of point sources. Alternatively, we gave zero weight
to pixels with strong point sources and averaged over the rest.
Both methods yielded very similar results.

The power in the harmonic
components from $m=1$ to $m=6$ is shown in Figure 10 for the K band image
as a function of radius.
The power in $m=1$ signifies the lopsidedness of the disk,
presumably due to the interaction with NGC~5195.
Aside from $m=1$  most
of the power is in the $m=2$ component, reflecting the strong two arm spiral.
The local maximum at $R=15\asec$ is the imprint of the central
bar onto the $m=2$ component.
Significant, but much less, power is also found in the $m=4$ and $m=6$
components, because the two spiral arms are more localized than a simple
double sinusoidal wave, i.e. the innerarm regions are wider
than the arms. There is only little power in the higher order components
with odd symmetry. There is also very little power in even components
higher than $m=6$; hence a Fourier decomposition up to $m=6$ yields a good
representation of the galaxy image.

Even though we have argued above on the basis of the $K$-$CO(2.3\mu )$
comparison that the K band amplitude is not significantly affected
by population changes, it is useful to employ the multi-color
information to check this.
Figure 11 compares the radial dependence of the $m=2$ component
at K and at V. The qualitative, global amplitude
structure is common to both wave bands,
showing two pronounced minima at $45\asec$ and $170\asec$ and a maximum
at $120\asec$. At $70\asec $ and $120\asec $, the optical colors exhibit an
amplitude which is 30\% lower than at $K$. These are the radii
at which the dust lanes, which are optically thick at all
wavelengths except $K$, are most extensive. This difference suggests
that at least one third of the flux in the spiral crest is absorbed by dust
in the optical bands, consistent with the optical depth estimates
from Section 3.3.

To obtain some estimate for the peak-to-valley amplitude of the two arm spiral
density contrast
we must use more information than the $m=2$ amplitude, because the spiral arms
are more peaked than a doubly sinusoidal curve. In Figure 13 we show
the sum of all {\it even} components up to $m=8$ at four radii.
The K band amplitude rises from  $I_{max}/I_{min} \approx 1.8$ at
$R=70\asec$ to a factor of 3 at $R=125\asec$ and falling back to a factor
of 2 at $R=150\asec$. These values for the amplitude of the mass tracing
stellar population are a factor of 3 to 5 higher than estimated by
Schweizer(1976) for M51, but are a factor of 1.5 to 2 lower than
estimated by Elmegreen \et (1989). Furthermore, as is already evident from the
$m=2$ component,
the amplitudes at I and K are comparable with K being somewhat
higher; this finding is in contrast
with Wright \et 's (1991) claim that the K-Band
amplitude is a factor of two lower than the I band amplitude.

Elmegreen \et (1992) recently have presented
evidence that many spiral galaxies
have significant $m=3$ arms,
even though they may not be apparent in visual inspection.
In M51 they find evidence for three arm spirals between the radii of
$50\asec$ and $100\asec$. In Figure 11 we present multi-color information on
the
radial run of the power in the $m=3$ component. It becomes apparent from this
Figure
that the power in this component is strongly color-dependent, decreasing in
amplitude towards redder passbands. We are hence led to conclude that these
features
in M51 are not dynamical features, tracing mass, but rather features in gas and
dust only.

\medskip
\section{CONCLUSIONS}
\smallskip\par
The main conclusions of this paper are as follows:

\bigskip

\noindent$\bullet$  Quantitative IR surface photometry (J,K,CO) is possible
down to
surface brightnesses of $\mu_K\sim 20$mag, with an accuracy of $0.06$mag.
Differential measurements are accurate to 0.03mag. Thus even for face-on spiral
galaxies
IR-array detectors can provide high quality surface photometry out to radii
comparable
to $R_{25}$.

\bigskip

\noindent$\bullet$ Even though the optical colors of face-on disks
are surprisingly insensitive to a smooth distribution of
dust, we can place limits on the global optical depth of M51's disk of
$\langle \tau_V \rangle \ltorder 1$, if the dust has a radial
scale length comparable to the stars'. The
dust in M51 removes a significant fraction of emission
from the spiral arm crest at all wavelengths
$< 1\mu$; this causes the spiral arm amplitude
at K($2.2\mu$) to exceed the one at
I($0.8\mu$). The major dust lanes have an
optical depth of $\tau_K\sim 0.4$,
indicating an amount of dust in the arms compatible with the amount
of molecular gas found there.
The relative scaleheights of dust and stars, $\zeta$,
is found to be $\sim 0.2-0.6$.
The fraction of radiation at $K(2.2\mu)$ removed by the dust
does not exceed $\sim 10\%$, even in the main dust lanes.

\noindent$\bullet$ We were able to show from measuring the CO$(2.3\mu )$
molecular
band strength across spiral arms that in localized regions young, red
supergiants
contribute significantly to the K-band flux. Since these regions are found to
be small,
this provides the first well founded evidence that even in
actively star forming galaxies K band images
{\it are indeed a good tracers of the massive, older disk population}.
Hence, variations in the K-band can be straightforwardly interpreted as
variations
in the stellar disk mass density.

\noindent$\bullet$ The spiral arm amplitude in M51,
as reflected in the K band image, is large:
up to a factor $\gtorder 3$. This finding is intermediate
to previous claims which were discrepant
by an order of magnitude.
The spiral arm amplitude exhibits well defined radial
variations with two pronounced minima at  $45\asec$ and
$170\asec$.

\noindent$\bullet$ We find that although the two arm spiral is clearly a mass
feature,
the reported features with threefold symmetry are color dependent, nearly
disappearing
at K. Hence, features with m=3 are most likely to be
present only in the dust and the gas, but not in the stars.

This and other imaging studies of spiral galaxies with infra-red array
detectors show
that spiral galaxies look quite different in the IR ($2.2\mu$) than at
I($0.8\mu$),
the longest wavelength at which they had been studied in detail so far. In
particular small bars are
common and the spiral arms seem to be strong. With imaging at K there finally
is a direct
way to map the stellar mass distribution in spiral galaxies in detail,
relatively
unafflicted by dust and young stars. Clearly, M51 with its peculiar interaction
history
does not allow generalizations to the properties of isolated spiral galaxies,
but only
enables a direct comparison with numerical simulations (e.g. Hernquist 1990).
Nevertheless, it is a good test case to establish the usefulness of IR surface
photometry
to map the dynamical structure of the galaxy. Ongoing studies of samples of
spiral galaxies will
allow a more complete survey for bars, will allow a check on how the spiral
arm amplitude  correlates with the current star formation rate and
will show how density wave resonances are reflected in the radial
variations of the spiral arm amplitude.

HWR was supported by a Hubble fellowship grant (HF-1024.01-91A)
and financial support for the NICMOS2 and 3 cameras
was provided by the National Science Foundation.

\medskip
\centerline{REFERENCES}
\smallskip\par

\Rf{Aaronson,~M., 1977, Ph.D. Thesis, Harvard University}

\Rf{Boroson~T., Strom,~K. and Strom,~S., 1983, ApJ, 274, 39}

\Rf{Bothun,~G., Mould,~J., Schommer,~B. and Aaronson,~M., 1985, ApJ 291, 586}

\Rf{Bruzual~G.~A., Magris~G.~C. and Calvet,~N., 1988, ApJ 333, 673}

\Rf{de~Vaucouleurs,~G., de~Vaucouleurs,~A. and Corwin,~H., 1976, Second
Reference Catalog of
Galaxies, Austin, Univ. of Texas Press}

\Rf{Disney,~M., Davies,~J. and Phillips,~S., 1989, MNRAS, 239, 939}

\Rf{Draine,~B.~T. and Lee,~H.~M., 1984, ApJ, 285, 89}

\Rf{Elmegreen,~D., 1980, ApJS, 43, 37}

\Rf{Elmegreen,~B., Elmegreen,~D. and Seiden,~P., 1989, ApJ, 343, 602}

\Rf{Elmegreen,~B., Elmegreen,~D. and Montenegro,~L., 1992, ApJS, 79, 37}

\Rf{Frogel,~J., Persson,~S., Aaronson,~M. and Matthews,~K., 1978, ApJ, 220, 75}

\Rf{Hernquist,~L., 1990, in "Dyn. and Interactions of Galaxies", ed. Wielen,
Springer, p.~108}

\Rf{Kaufman,~M., Bash,~F., Hine,~B., Rots,~A.,
Elmegreen,~D. and Hodge,~W., 1989, ApJ, 345, 674}

\Rf{Kylafis,~N. and Bahcall,~J.~N., 1987, 317, 637}

\Rf{Peletier,~R. and Willner,~S., 1991, ApJ, 382, 382}

\Rf{Peletier,~R. and Willner,~S., 1992, AJ, 103, 1761}

\Rf{Peletier,~R., 1989, Ph.D. Thesis, University of Groningen}

\Rf{Persson,~S., Frogel,~J., and Aaronson,~M. 1979, ApJS, 39, 61}

\Rf{Rand,~R. and Kulkarni,~S., 1990, ApJL, 349, L43}

\Rf{Rieke,~G., Loken,~K., Rieke,~M. and Tamblyn,~P. 1993, ApJ, in press}

\Rf{Schild,~R., 1983, AJ, 95, 1021}

\Rf{Schweizer,~F.,1976, ApJS, 31, 313}

\Rf{Toomre,~A. and Toomre,~J., 1972, ApJ., 178, 623}

\Rf{Tully,~R., 1974, ApJS, 27, 449}

\Rf{Valentijn, E.~A., 1990, Nature, 346, 153}

\Rf{Wainscoat,~R., Hyland,~A. and Freeman,~K., 1990, ApJ, 348, 85}

\Rf{Witt,A.~N., Thronson,~H.~A. and Capuano,~J.~M., 1992, ApJ, 393, 611}

\Rf{Wright,~G., Casali,~M. and Walther,~D., 1991, in "Astrophysics with IR
Arrays",
ed. R. Elston, Cambridge Univ. Press}

\medskip
\centerline{Figure Captions}
\smallskip\par

\medskip
\noindent{\bf Figure 1 : } $7.2^{'}\times 7.3^{'}$ images of M51
in B~V~R~I and K.

The K band image is a mosaic of five images. West is up and
North is to the right.
The pixel scale is $1.8\asec /pix$ as indicated by the $60\asec$ bars
in the B band image. The optical images
have been binned to the same resolution as the K band image.
Note that in all bands shortward of $1\mu$ the major dust
lanes associated with the spiral arms are still prominent.
The impact of dust only becomes small at K.

\medskip
\noindent{\bf Figure 2 : } $2.5^{'}\times 2.5^{'}$ images of M51
in V,~J,~K,~CO$(2.3\mu )$ and $J-K$.

Images of the central region of M51 in V,~J,~K and CO$(2.3\mu )$
taken with the NICMOS3 camera at $0.6\asec /pix$. Orientation
and scale of the image are indicated in V band image. The K band
image shows very clearly the central bar. The J--K color image
outlines the  position of the strongest dust lanes.

\medskip
\noindent{\bf Figure 3 : } Calibration of the IR Photometry.

We compare the cumulative magnitudes in J and K (as measured from
the surface photometry presented here) with the aperture
measurements by Aaronson (1975), shown as triangles.
The dashed line represents the data from the $7^{\prime}\times 7^{\prime}$
mosaic; the solid line represents the high resolution
data. The solid line is continued as a dotted line for radii
where the data do not provide full azimuthal coverage
(see Figure 2). For the two apertures the
agreement is excellent ($0.03$mag), and we use
Aaronson's measurements for the absolute flux calibration.

\medskip
\noindent{\bf Figure 4 : } Radial Luminosity Profiles of M~51
in B~V~R~I~J and K.

This figure shows the radial luminosity profiles in all bands measured.
The flux in each band was sampled in concentric annuli, after
removing point sources. The error bars shown include both statistical
errors and the uncertainty in the sky level. Only the central part
of M51 was imaged in J, hence these data have a smaller radial extent.
The K band profile from the mosaic with the NICMOS2 camera and
the high resolution data with the NICMOS3 camera are overplotted
as the top profile.

\medskip
\noindent{\bf Figure 5 : } Radial Color Profiles of M~51
with Respect to I.

This figure shows the radial color profiles with respect to I
for all measured bands. The dash dotted line indicates the
approximate radius of the transition from the bulge to the disk.
The optical data are only shown to radii to which reliable
K band surface photometry was obtained.

\medskip
\noindent{\bf Figure 6 : } Model
for Color Change in the Presence of Dust.

This figure shows the
expected color change, with respect to the unreddened case,
for the star dust model described in the Appendix. The model assumes
Gaussian stratifications of the dust and the stars, with a
scale height ratio of $\zeta$ and a total optical depth
of $\tau_V$. Models with $\zeta=0.4$ are shown for two
inclinations: nearly face-on (cos(i)=0.9) and highly inclined
(cos(i)=0.3). It is noteworthy how little ($\ltorder 0.1$mag) the colors
change for the face on case even as $\tau_V$ approaches unity.
This is due to the high dust albedo in the optical which makes
the dust layer reflective to radiation from in front of it.

\medskip
\noindent{\bf Figure 7 : } Color Cut Across Dust Lanes.

This figure shows the color change across two dust lanes
with respect to K in all other bands. The top panel shows a dust
lane in the bulge/disk transition region ($\sim 25\asec$)
and the bottom panel shows a lane in one of the dominant
arms at $R\sim 90\asec$. As expected, the color changes
become monotonically stronger towards bluer colors
(J-K, I-K, R-K, V-K and B-K).

\medskip
\noindent{\bf Figure 8 : } Confidence Limits in the
$\tau_V-\zeta$ plane for Two Dust Lanes.

This figure shows confidence limits in the $\tau_K-\zeta$
plane when applying the models from the Appendix to the
color changes shown in Figure 7. The contours are spaced by
$\Delta\chi^2=3$ compared to the best model. For the dust lane at $25\asec$
the optical depth at K is found to be $\sim 0.4$ (corresponding
to $\tau_V\sim 3.5$) with a scale height
of 60\% of the star's. For the dust lane in the disk (lower panel)
the dust scale height is only $\sim35\%$ of the star's, and the optical
depth is constrained to be $\tau_K=0.4-1$, corresponding to
$\tau_V=3.5-9$.

\medskip
\noindent{\bf Figure 9 : } K--CO$(2.3\mu$)change across Spiral
Arms.

This figures compares the K band to the CO$(2.3\mu$) narrow
band profiles along a cut that crosses two spiral arms.
Since the depth of the CO$(2.3\mu$) is a strong function
of surface gravity, a dip in the CO$(2.3\mu$) profile,
compared to K, is expected if much of the K light
comes from young, red supergiants. This cut shows the only
place in the inner $2.5^\prime\times 2.5^\prime$ where supergiants
appear to be important contributors
to the K band light, just outside the crest of the spiral arm
(around pixels 160).
Below we compare the K and J light profiles along the
same cut. This comparison shows a smooth color change between arm and
inter-arm region of ($\sim 0.1$mag) and the main dust lane (near pixel 150).

\medskip
\noindent{\bf Figure 10 : } Harmonic Decomposition of the Spiral
Arms at K.

This figure shows the result of a Fourier decomposition
in azimuth of the intensity in concentric annuli of the K band image.
The K band image had been corrected to face-on.
The panels show the
$m=1...6$ Fourier amplitudes, normalized by the mean flux in each
annulus. There is a non-zero $m=1$ component at all radii,
indicating the lopsidedness of M51. The dominant component
is $m=2$, arising in the inner parts $R\ltorder 25\asec$ from
the central bar and in the outer parts from the strong
two arm spiral. Note the clear minima in the radial run of the
$m=2$ amplitude at $45\asec$ and $170\asec$.
There is significant power in the $m=4$ component, because the two
arm spiral is a more compact feature than a doubly sinusoidal wave.
The power in all odd components, except $m=1$ is small
($\ltorder 10\%$).

\medskip
\noindent{\bf Figure 11 : } K--V Comparison of the Harmonic
Decomposition.

This figure compares the $m=2$ and the $m=3$
azimuthal components in V and K. For the $m=2$ component
the V and K amplitudes trace each other at least qualitatively.
Note that at some radii the K band  image has about 25\%
more $m=2$ power, because the signal is suppressed by dust in the
optical. Thus the two arm spirals are more prominent in
K than in the optical.
Even though the V band seems to show significant $m=3$ power
at $\sim 70\asec$, such a feature is not present in K.
Thus there is no massive 3-armed spiral in M51.

\medskip
\noindent{\bf Figure 12 : } Arm--Interarm Density Contrast.

This figure
shows the symmetrized azimuthal brightness variations (in B~I and K)
at four different radii, obtained by adding all even Fourier
components up to $m=8$. The solid line represents the K band variation
and is expected to be a good approximation (see Section 4.3.2) to the
surface mass density variations. The I band profiles
(dashed lines) already show the significant impact of the dust lane
(to the right of the crest). The B band profile (dotted line) shows both
the impact of dust as well as the young blue stars on the trailing
(left side) side of the crest.
This figure shows that the arm-interarm contrast in K is as large
as 3.

\page
\centerline{\bf APPENDIX}
\smallskip\par

In this appendix we briefly describe the radiative transfer calculation used to
estimate the expected color changes due to dust. The treatment here is very
similar
to the one described by Elmegreen (1980). We assume that both the stars and the
dust are stratified in plane-parallel geometry.
The vertical density profile is
assumed to be a Gaussian for both components, with scale heights $\beta_{star}$
and $\beta_{dust}$:
$$\kappa (z) = \kappa_0~exp\Bigl(-{ {z^2}\over{2\beta^2_{dust}}
}\Bigr)$$
$$j (z) = j_0~exp\Bigl (-{ {z^2}\over{2\beta^2_{stars}} }\Bigr),$$
where $j$ is the volume emissivity of the stars and $\kappa$ is the total
cross section per unit length of the dust.  Both $j$ and $\kappa$ are of course
functions of wavelength.
To calculate colors with respect to the unreddened case, it is sufficient to
specify each
model by $\zeta\equiv \beta_{dust}/\beta_{stars}$ and by the total optical
depth
at a fiducial wavelength
$$\tau\equiv\int_{-\infty}^{\infty}\kappa (z)dz =
\sqrt{2\pi}\kappa_0\beta_{dust}.$$

We only include absorption and isotropic scattering
into the dust cross section,
$\kappa = \kappa_{abs} + \sigma_{iso}$, with values
for various wavelengths given by Draine and Lee (1984).
The (isotropic) albedo is denoted as $\omega\equiv
{{\sigma_{iso}}\over{\kappa}}$.
As a further simplification we treat single scattering only
and neglect multiple scattering.
This implies that
$$ I(\mu , z^{'}) = {1\over \mu }\int_{-\infty }^{z^{'}}
j(z^{''})exp({{\tau^{''}-\tau^{'}}\over\mu })dz^{''}$$
for $\mu > 0$ (where $\mu=cos(\theta )$ and $\theta$ is the angle with respect
to the normal vector
of the stratification). For $\mu<0$, $I(\mu,z^{'})= - I(-\mu,-z^{'})$.

The mean intensity of $J(z)$ at a height $z$ is defined as
$$J(z^{'})= 2\pi\int_{-1}^{1}I(\mu^{'},z^{'})d\mu^{'}$$.

We can now calculate for each wavelength and each optical depth
the intensity emerging from the surface at an angle $\mu$ as
$$I_{\infty }(\mu )={1\over \mu }\left [ \int_{-\infty}^{\infty}
\left [ j(z^{'})+{\omega\over 2}\kappa(z^{'})J(z^{'})\right ]
\cdot exp\Bigl({{\tau^{'}-\tau}\over\mu }\Bigr)dz^{'}\right ].$$
This allows us to calculate the color change (Section 4.1) in the observed
bands
(with respect to the unobscured case) for a disk
seen at inclination $\mu$ for various optical depths, specified by, say,
$\tau_V$.
Alternatively, we can for each wavelength calculate the expected flux
attenuation
across a dust lane (Section 4.2) with optical depth, $\tau$ (assuming that the
optical depth
outside the lane is much smaller).
\vfill\eject
\begin{center}
\begin{tabular}{c c c}
&TABLE 1&\\
\noalign{\bigskip\hrule\smallskip\hrule\medskip}
Color& Lane 86& Lane 45\\
$\Delta C$&[mag]&[mag]\\
\noalign{\medskip\hrule\medskip}
 J-K&$0.21\pm0.04$&$0.13\pm0.04$\\
 I-K&$0.50\pm0.04$&$0.38\pm0.04$\\
 R-K&$0.64\pm0.04$&$0.53\pm0.04$\\
 V-K&$0.79\pm0.04$&$0.58\pm0.04$\\
 B-K&$1.01\pm0.04$&$0.69\pm0.04$\\
\noalign{\medskip\hrule\medskip}
\end{tabular}
\end{center}
\end{document}